\documentclass[%
 amsmath,amssymb,
 aps,
 prl,twocolumn,showpacs
]{revtex4-1}

\usepackage{graphicx}
\usepackage{dcolumn}
\usepackage{bm}
\usepackage{hyperref}
\usepackage{color}

\begin{document}
\preprint{Physical Review Letters}

\title{Topology, geometry and mechanics of  surgical Z-plasty}

\author{Elisabetta A. Matsumoto$^{1,2}$, Haiyi Liang$^3$, L. Mahadevan$^{1,4}$}
\thanks{The first two authors contributed equally} \email{Author for correspondence: lmahadev@g.harvard.edu}\affiliation{$^1$ Paulson School of Engineering and Applied Sciences,\\Wyss Institute for Biologically Inspired Engineering, \\$^2$ School of Physics, Georgia Institute of Technology, 837 State Street, Atlanta, GA 30332, USA \\$^3$ CAS Key Laboratory of Mechanical Behavior and Design of Materials, University of Science and Technology of China, Hefei, China,\\ $^{4}$ Department of Physics and Kavli Institute for NanoBio Science and Technology, Harvard University, 29 Oxford Street, Cambridge, MA 02138, USA}
 \date{\today}

\begin{abstract}

Reconstructive surgeries often use topological manipulation of tissue to minimize post-operative scarring.  The most common version of this, Z-plasty, involves modifying a straight line cut into a Z-shape, followed by a rotational transposition of the resulting triangular pedicle flaps, and a final restitching the wound. This locally  reorients the anisotropic stress field and reduces the potential for scarring. We analyze the planar geometry and mechanics of the Z-plasty to quantify the rotation of the overall stress field and the local forces on the restitched cut using theory, simulations and simple physical Z-plasty experiments with foam sheets that corroborate each other. Our study rationalizes  the most typical surgical choice of this angle, and opens the way for a range of surgical decisions by characterizing the stresses along the cut.

\end{abstract}

\maketitle

The wound healing response of damaged tissue is a complex process associated with fibroblast proliferation and movement, extracellular matrix (ECM) synthesis, and leads to tissue remodeling \cite{Barrandon}. The physiology of wound healing depends on the local mechanical environment of the wound and the remodeled skin frequently has different elastic properties than healthy skin, resulting in inhomogeneous deformations  \cite{Cerda2005}. In severe cases, hypertrophic scaring develops as skin fibroses after surgical or traumatic injuries, and requires surgical intervention to restore function and {\ae}sthetics, and include a suite of flap based ``plasty" surgeries \cite{Limberg1984,Grabb2008}.  The most basic of these plasties is the Z-plasty, first introduced more than 175 years ago to correct the coarctation of the eyelid \cite{Horner1837,Borges1973}. It involves a topological transformation of a ``Z"-shaped incision that exposes two triangular flaps. When these flaps are transposed and stitched together, the original linear cut (the central limb of the Z) rotates, reducing the opening stress on the wound (Fig. 1a) \cite{figure}. This prevents the scarification of linear wounds by approximately reorienting the cuts parallel to the local directions of maximal tension in the skin \cite{Perez2013} which minimizes fibrosis, and by anisotropically lengthening tissue also serves to restore mobility. Here we present an elastic model for Z-plasty with the aim of understanding both the geometry and stresses in the final structure introduced by this topological transformation, with the aim of rationalizing the surgical design rules for the optimal angle of the additional incisions.

\begin{figure}\label{fig1}
\includegraphics[width=0.35\textwidth]{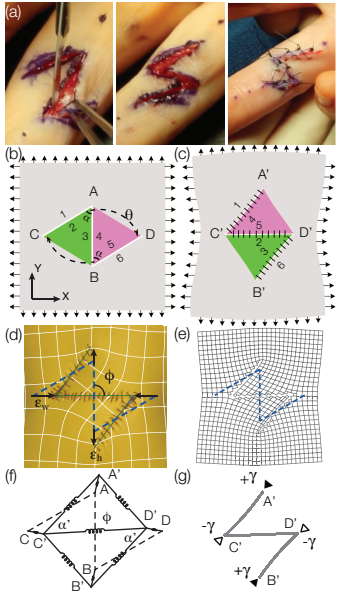}
\caption{(a) A Z-plasty procedure is used to release the (burn-induced scar) contraction of a finger joint by relaxing the strains induced by scarring (from \cite{figure}) (b) The cuts that characterize the Z-plasty; AB is the original cut while the new cuts AC and BD make an angle $\alpha$ with the original one and have a length $\ell$. (c) The topological reconnection.  The sides that are originally paired, i.e. 1-2, 3-4, 5-6 are rotated and stitched together as 1-4, 2-5 and 3-6 after surgery. This causes the flaps to rotate by an amount $\theta$ and the limbs to rotate by an amount $\phi$.  (d) A z-plasty in a foam-rubber sheet shows the strong local distortion of the mesh. This geometric transformation is accompanied by a lateral shrinkage strain $\varepsilon_w$ and a transverse extension strain $\varepsilon_h$. (e) Numerical simulations of an elastic sheet subject to Z-plasty show a distorted mesh around a Z-plasty. (The dashed lines outline the initial incision.) (f) A  coarse-grained approximation with 5 discrete springs captures the essence of the transformation. (g) At a continuum level,  Z-plasty is  a disclination quadrupole made of two positive disclinations of strength $\gamma$ and two negative disclinations of strength $-\gamma$ that lead to local compression, stretching and shear.}\end{figure}

Our starting point is skin that is subject to a uniaxial prestress field, mimicking the tension along Langer's lines in the skin.  In Z-plasty, a primary wound of length $\ell$ (chosen here to be oriented perpendicular to this principal stress $\bf e_y$) \cite{footnote1}, depicted as AB in Fig. 1b, is made seemingly worse by introducing two auxiliary incisions at either end of the wound, AC and BD, at an angle $\alpha$ to the AB \cite{footnote3}.  The result is the appearance of two similar triangular flaps, BCA and ADB, whose shared middle limb corresponds to the primary incision (Fig. 1b).   Next, the two flaps are excised from the underlying tissues, transposed and stitched together to yield a new topological configuration (Fig. 1c).  The resulting ``S"-shaped structure has a middle branch created by joining two flaps along their auxiliary incisions; when this structure is allowed to reach elastic equilibrium, the flaps rotate by an amount $\theta$ (Fig. 1b) while central limb of the Z, which corresponds to the original wound, rotates by an amount $\phi$ (Fig. 1d). Both these angles are functions of the cutting $\alpha$ and the ambient stress field. This geometrical transformation is accompanied by a transverse lengthening strain $\varepsilon_h$ and a lateral shortening strain $\varepsilon_w$.   Mimicking the basic Z-plasty surgery on a sheet of foam-rubber, as shown in Fig.1d, we see that a reference mesh in the vicinity of the ``wound" is strongly distorted locally, but remains relatively undistorted in the far field. 

\begin{figure}\label{fig2} 
\includegraphics[width=0.35\textwidth]{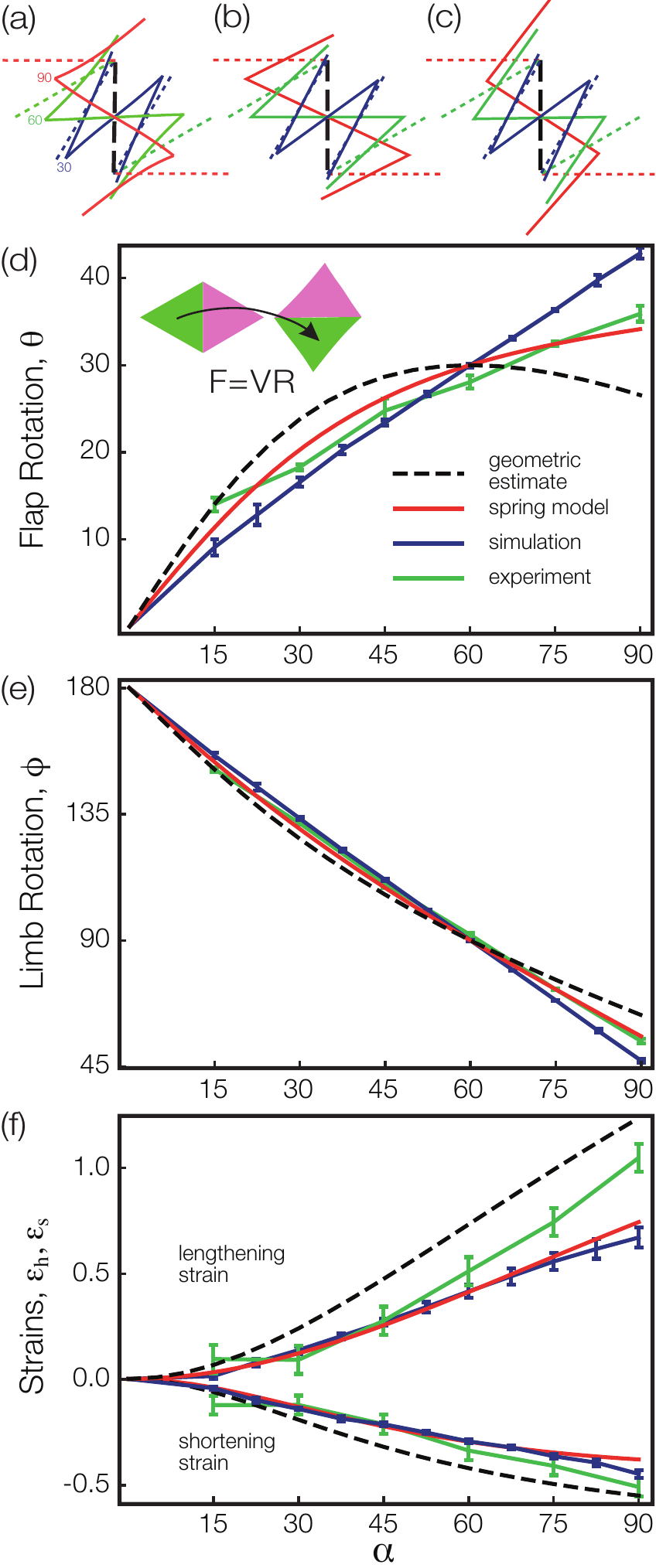}
\caption{
a-c) Initial (dotted lines) and final (solid lines) geometries for different incision angles $\alpha=30^\circ, 60^\circ, 90^\circ$ for pre-strain $0.5:0.25$.  Results from the fine-grained simulation are shown in a), and those from the simple spring model in b), while in c), we show the results from a simple physical experiment using a sheet of foam rubber. The rotation angle of d) a flap, $\theta$, e) the central limb, $\phi$ as a function of the incision angle $\alpha$. f) Transverse lengthening $\varepsilon_h$ and lateral shortening $\varepsilon_w$ as a function of incision angle $\alpha$. }
\end{figure}

Surgeons typically estimate the final shape of a Z-plasty by ignoring any contributions from elastic deformation and assuming rigid rotation of the parallelogram flaps so that the long diagonal is aligned along ${\bf e}_y$ after the transposal \cite{McGregor1957,Ellur2009}. Geometry alone dictates that the rigid rotation of the left flap from $\Delta CAB$ to $\Delta C'D'B'$ is $\theta^g=\sin^{-1}(\sin\alpha/\sqrt{5-4\cos\alpha})$, where the superscript $g$ denotes surgeon's geometric estimate. The rotation of the central limb counterclockwise with respect to the $+{\bf e}_y$ direction is $\phi^g=\cos^{-1}(\vec{A}\cdot\vec{D}'/(|A||D'|))=\cos^{-1}((1-2\cos\alpha)/\sqrt{5-4\cos\alpha})=180-\alpha-\theta^g.$    Then the lengthening along $\bf Y$ axis and the shortening along the $\bf X$ axis correspond to the difference between the long diagonal $\overline{CD}$ and the short diagonal $\overline{AB}$, and yield the lengthening strain defined as $\varepsilon^g_h=(5-4\cos \alpha)^{1/2}-1$ and complementary shortening strain, $\varepsilon^g_w=-\varepsilon^g_h/(1+\varepsilon^g_h)$. In the case of the basic Z-plasty \cite{McGregor1957,Ellur2009}, surgeons typically assume  $\alpha=60^\circ$, and further that the length of the  incision is the same as that of the original cut. The simple geometric estimate predicts that the central limb rotates clockwise by $90^\circ$ from ${\bf e}_y$ to ${\bf e}_x$, the height increases by $73\%$, and the width decreases by $42\%$. This also corresponds to the maximal possible rotation. Such purely geometric estimates ignore the elastic response of skin \cite{Furnas1971}, and raise the question of how these would effect the final state.  

We consider two closely related elastic models for the behavior of skin: the first a finely discretized mesh of springs (Fig. 1e) and the other represents the flaps and the incisions using coarse-grained springs (Fig.1f).  In either case, we see that the final outcome of a Z-plasty is equivalent to introducing a disclination quadrupole of strength $\alpha$ at the wound site (Fig. 1g). We note that although a single disclination has unbounded deformation and divergent elastic energy density that scales as $~r^2$ and a dipolar deformation has an elastic energy density that grows as $\ln r$, a disclination quadrupole has zero net disclination charge and restricts the deformation to a finite region\cite{Romanov2003}. Furthermore, as it only involves the coarctation of triangles, the quadrupole is the lowest order disclination multipole that preserves the original tissue area, minimizes deformations in the far field and has an energy density that decays as $~r^{-2}$ \cite{Nabarro, Matsumoto2009}, making Z-plasty practical. It is interesting to note that these facts were known practically to surgeons a long time before dislocations and disclinations were formally introduced in physics \cite{Nabarro}. However, in the case that we consider here, the elastic deformation around the quadruple involves both large rotation and large strain, making the problem more difficult than the linear theory of elastic dislocations.

Our minimal spring model ignores the continuous deformation of the flaps and the surrounding regions and instead treats just the displacements of the vertices $\{A,B,C,D\},$ to their new locations $\{A',B',C',D'\},$ and  the deformations of each side of the incisions which are treated as spring-like edges, with natural lengths corresponding to the lengths of the initial incisions. Because of the initial symmetry, we shall assume that $\overline{A'C'}=\overline{B'D'},$ $\overline{AA'}=\overline{BB'},$ $\overline{CC'}=\overline{DD'}$ and $\angle A'C'D'=\angle C'D'B'$.  This results in the energy:
\begin{eqnarray}
&F&=\textstyle\kappa\left(\overline{AA'}\right)^2+\kappa\left(\overline{CC'}\right)^2+\kappa\left(\overline{C'D'}-1\right)^2\nonumber\\
&&\textstyle+\kappa\left(\overline{A'C'}-1\right)^2+\kappa\left(\overline{A'D'}-2\sin(\frac{\alpha}{2})\right)^2,
\end{eqnarray}
where we have assumed that all springs have an identical elastic constant $\kappa$. Setting the values of $A'=\{x_a,y_a\}$ and $C'=\{x_c,y_c\}$, we can numerically minimize the elastic energy for different value of $\alpha$ to find a rough estimate of the geometry of the incisions post Z-plasty.  

Our more refined model treats skin via a discretized equilateral triangular mesh of freely-jointed linear elastic springs where the energies and lengths have been rescaled such that the spring constant and rest length are unity;  this simple network yields results similar to those obtained using finite element simulations \cite{Sifakis2009,Kitta2013}, but is inexpensive. The spring network naturally accounts for geometric nonlinearities arising from large deformations. We use a mesh of size $500\times500$, and first pre-stretch the sheet anisotropically with traction forces $\textbf{f}_0$ applied at the boundary nodes so that $\epsilon_x^0 > \epsilon_y^0$ . To mimic the process of Z-plasty surgery, we use the following protocol: (1) Cutting- The linear z-shaped incision is generated on the pre-stretched mesh by breaking springs along the incision. (2) Topological transposition- A geometric transformation $\textbf{x}=\textbf{F}_0 \textbf{X}$ is applied to the flaps so that $\Delta CAB\rightarrow\Delta CDB$ and $\Delta ABD\rightarrow\Delta ACD$  (shown in the inset of Fig. 2b), where $\textbf{x}$ and $\textbf{X}$ are the new and old positions of the flap, and $\textbf{F}_0$ is the deformation gradient defined by the three vertices of a flap before and after transposition \cite{Belytschko2000}. (3) Stitching and relaxation- The cut between the relocated limbs is ``stitched" by adding new springs connecting opposing nodes, and  the elastic energy of the spring network is minimized via a conjugate gradient method with constant traction $\textbf{f}_0$ at the boundaries. As a result, the relocated flaps induces a strong local distortion and displays lengthening and shortening, as shown in Fig. 1c),d). We quantify the mechanical response for different tip angles $\alpha \in [15^\circ,90^\circ]$, limb length $\ell=100$, Poisson ratio $\nu=0.3$ \cite{Choi2005} and different pre-strain $\varepsilon^0_x:\varepsilon^0_y=0.4:0.2,\, 0.5:0.25,\, 0.6:0.2,\, 0.6:0.3, \, 0.6:0.4, \, 0.6:0.5.$

In Fig. 2a-c) we show the results from both these models for  three angles $\alpha=30^\circ, 60^\circ, 90^\circ$ subjected to the same pre-strain $\varepsilon^0_x:\varepsilon^0_y=0.5 : 0.25$: numerical simulations (Fig. 2a), simple spring model (Fig. 2b) and compare them with the results of our physical experiments carried out using foam sheets (Fig 2c).  We note that since the strains are not always small, we need to use to a correctly invariant form of the strain - we use discrete analogs of the Green-Lagrange strain here based on the deformation gradient, i.e. the components of $\mathbf{E} = \tfrac{1}{2} (\mathbf{C} - \mathbf{I})$, where $\mathbf{C} = \mathbf{F}^T\mathbf{F}$ is the right Cauchy-Green deformation tensor \cite{Marsden, Ogden, Bonet, Belytschko2000}. Four quantities serve as the basis for our comparison between the geometric (surgical) estimate and our two models: the rotation of each flap, $\theta$ (Fig. 2d), the rotation of the limb $\phi$ (Fig. 2e) and the strains, $\varepsilon_h$ and $\varepsilon_w$ (Fig. 2f)\cite{footnote2}. The transformed geometries based on both our elastic models agree qualitatively with one another for small angles, but differ from each other when $\alpha$ is large. They also differ substantially from the geometric model. In the purely geometric model, the flaps are not allowed to shear, and the final triangular flap is congruent to the initial one. Both spring models allow for the elastic deformation to be partially accommodated by a shearing transformation, which accounts for the qualitative difference from the geometric model.  Most interestingly, our physical experiment and all three models predict the same limb rotation angle of $90^\circ$ for just one value of the incision, $\alpha =60^\circ$, and thus potentially minimize tearing stress on an anisotropically loaded wound. This result provides a physical justification for the commonly used surgical choice of $\alpha=60^\circ$ Z-plasty, as it is the only value where the geometric and elastic models coincide. In Fig. 2f), we show the lengthening strains perpendicular to the original cut $\epsilon_h$ and shortening strain parallel to it $\epsilon_w$, and again see that while the elastic models are in qualitative agreement, the geometric estimate overestimates the post-plasty strain; the source of this discrepancy lies in the strong local nonlinearity of the deformation field near the disclinations. We note that the  deviation between the models and experiment occurs in the lengthening strain $\varepsilon_h$ for very large $\alpha$, which corresponds to the strongly non-linear regime; in this case, the spring model provides results that are surprising close to our experiments, unlike the detailed simulations.  

\begin{figure}\label{fig3} 
\includegraphics[width=0.35\textwidth]{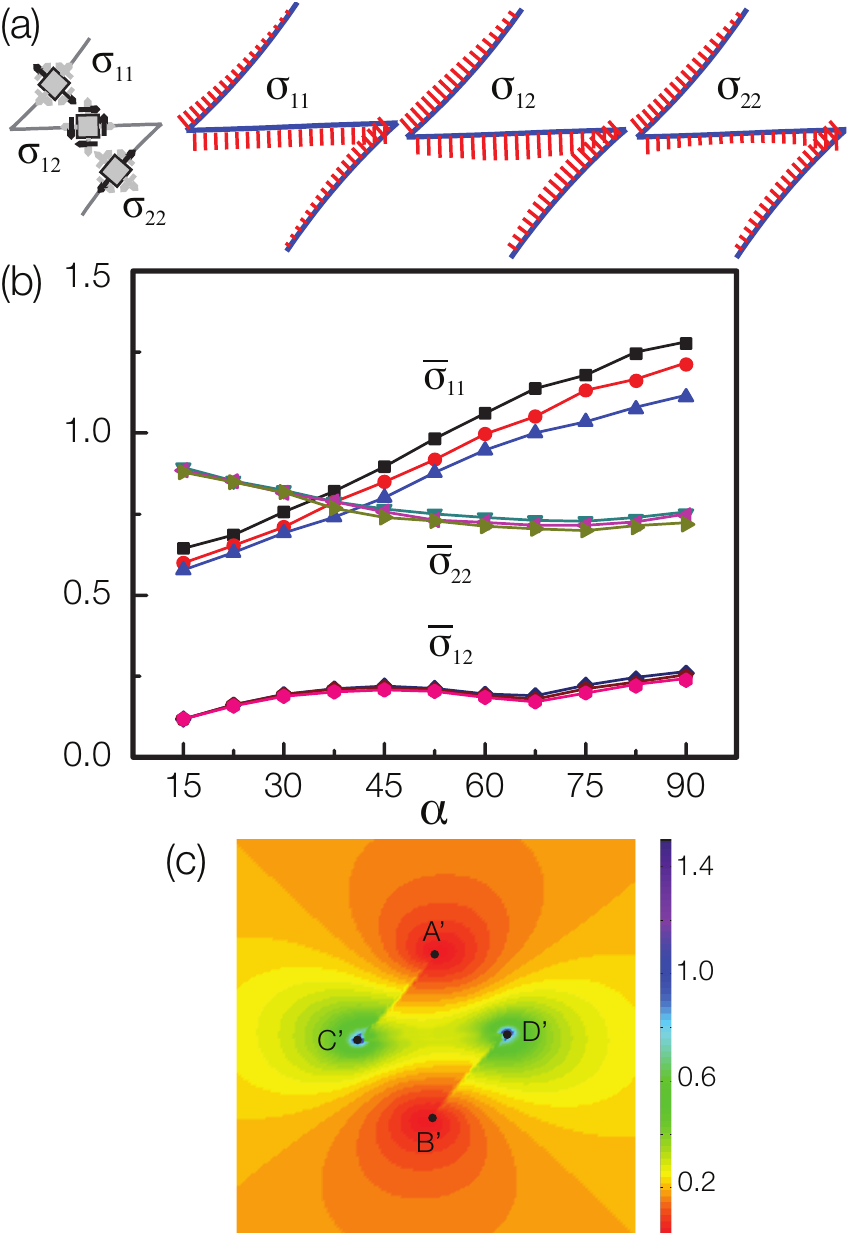}
\caption{The Cauchy stress and elastic energy obtained from our refined simulation. a) The definition of crack-opening stress $\sigma_{11}$, shearing stress $\sigma_{12}$, and crack-buckling stress $\sigma_{22}$ along an incision. b) The averaged stress line integrals of $\overline{\sigma}_{11}$, $\overline{\sigma}_{12}$, and $\overline{\sigma}_{22}$ as function of tip angle $\alpha$. The different curves correspond to different values of the pre-strain ration $\varepsilon_x : \varepsilon_y$, black $0.6 : 0.2$, red $0.6 : 0.3$, blue $0.6 : 0.4$,  c) The elastic energy field with pre-strain $\varepsilon^0_x:\varepsilon^0_y=0.5 : 0.25$ and tip angle $60^\circ$,  corresponding to the mesh in Fig. 1e shows two maxima and two minima associated with the superposition of the disclination quadropole atop a uniform strain field.}
\end{figure}

Having evaluated the geometric consequences of an elastic model for the Z-plasty, we now evaluate the tension-relief effects of Z-plasty on the main wound and the auxiliary incisions. This is most naturally described in terms of the components of the (true) Cauchy stresses $\sigma_{11}$ associated with tearing open the cut, $\sigma_{12}$ shearing along the cuts, and $ \sigma_{22}$ compression along the cuts, shown in  Fig. 3a for a particular choice of $\alpha = 60^\circ$. Fig. 3b shows the average stress along the cut $C$,  defined by $\bar{\sigma}_{\alpha\beta}=\int_{C}\widetilde{\sigma}_{\alpha\beta} ds/\int_{C}ds$  rescaled by the horizontal pre-stress $\sigma_x^0$, for $15^\circ\leq\alpha \leq90^\circ$. We note that the average opening stress, $\overline{\sigma}_{11},$ increases monotonically with $\alpha$, while the other components are either constant or weakly decreasing with increasing $\alpha$. Noting the small change in the lengths of the side limbs, the main contribution to $\overline{\sigma}_{11}$ comes from the geometric lengthening of the central limb that scales with $\varepsilon_h^g=(5-4\cos \alpha)^{1/2}$.  The shear stress $\bar{\sigma}_{12}$ is relatively small, and does not vary much with $\alpha$. We are not sure how the monotonic increase in $\sigma_{11}$ with $\alpha$ is reflected in wound healing dynamics. Our averages reflect equal weighting across all wound, whereas damaged tissue near a wound site will be more susceptible to keloid formation due to the overall inflammatory response. In cases, where all the tissue is healthy, as in the hand procedure shown in Fig. 1(a), the goal is not reduction in scar formation, but an effective change in lengthening versus shortening strain in the area surrounding the incision. Whilst this produces an effective shear in the area, minimizing the shear-stress across the incisions will likely promote healing, and this is optimized at angles a little larger than 60° (Fig. 3b).
The energy density associated with the surgery contains four nearly singular regions associated with a disclination quadrupole, with two energetic maxima, at $C'$ and $D',$ and two minima, at $A'$ and $B'$, shown in Fig. 3c. 

Our study of a classical plastic surgery technique, the Z-plasty -- a topological rearrangement of skin in the neighborhood of a cut, complements previous geometrical approaches with an elastic analysis. Our elastic models compare reasonably with physical experiments but coincide with the geometric \emph{ansatz} used by surgeons only for $\alpha=60^\circ$, when limb rotation is maximized, allowing the primary wound to rotate by $90^\circ$. From a topological perspective, other surgical operations can be considered as multipole expansions of disclinations: e.g.  Y-plasty, which converts a ``V" shaped incision into a ``Y" shape leads to a small area of negative Gaussian curvature at the branch of the ``Y",  W-plasty and its generalization, wherein multiple Z-plasties are conducted along a long primary wound, which minimizes the lateral extent of the deformation whilst maintaining the local elastic benefits of Z-plasty. From a physical perspective, Z-plasty combines kirigami, where cuts are used to create distinct three-dimensional geometries \cite{Castle2014,Sussman2015} and elasticity, since the goal is to perform planar topological manipulations while keeping the final skin topography as close as possible to that of the underlying tissue.   From a biomedical perspective, our study sets the stage to account for the dynamics of healing as they are affected by the local strain state in the vicinity of the cut. From a surgical perspective, our model sets the stage for a consideration of stress fields and elastic deformations that surgeons can optimize when considering the initial choice of incision to increase efficacy of post-surgical outcomes, once we account for out-plane deformations as well, a natural next step.

{\bf Acknowledgments} We thank Amit Patel for contributing  to the project in its initial stages, the National Natural Science Foundation of China (11272303),  and the Aspen Center for Physics supported by NSF grant PHY-1607611 for their hospitality in the final stages of this work.


\begin{thebibliography}{9}
\bibitem{footnote1}{We chose the case of the initial wound direction being perpendicular to the stress direction as this is the worst case scenario, and the most difficult to deal with surgically as the wound has the tendency to gape maximally. In situations where the wound is not orthogonal to the pre-stress direction, there will be a component of the stress  acting along the wound; as long as this is not negative, it makes little qualitative difference to our results. If it is negative, there is a tendency for the cut to buckle, but as we consider only planar deformations, we do not consider this.}
\bibitem{footnote2}{Here we keep the lengths of the side limbs equal to the central wound, as it would otherwise lead to the effective insertion of dislocations and discontinuities in the traction along the stitched cuts.}
\end{thebibliography}
\end{document}